\documentstyle[baasar]{article}

\begin{document}

\title{\LARGE On the Importance of PhD Institute in Establishing a Long-Term
Research Career in Astronomy}

\author{Brad K.Gibson}
\affil{\vskip-0.8mm Center for Astrophysics \& Space Astronomy, Department of 
Astrophysical \& Planetary Sciences, University of Colorado,
Campus Box 389, Boulder, CO 80309-0389, bgibson@casa.colorado.edu}
\author{\vskip3.0mm Michelle Buxton, Emanuel Vassiliadis, Maartje N. Sevenster,
D.~Heath Jones}
\affil{\vskip-0.8mm Research School of Astronomy \& Astrophysics, Australian 
National University, Weston Creek P.~O., Weston, ACT 2611, Australia}
\author{\vskip3.0mm Rebecca K. Thornberry}
\affil{\vskip-0.8mm Orange High School, P.~O. Box 654, Orange, NSW 2800, 
Australia}

\begin{abstract}
We have examined the success rates of 
19 American, Canadian,
Australian, and Dutch graduate programs in producing long-term, career,
research astronomers.  A 20-year baseline was considered (1975-1994),
incorporating 897 astronomy PhD graduates.
The major conclusion from our study is that
the fraction of PhD graduates still involved in astronomical
research is surprisingly insensitive to the institutional source of one's PhD.
With few exception, $\sim 55\rightarrow 75$\% of astronomy graduates,
regardless of PhD source, remain active in the astronomical research community.
While it remains true that graduates of so-called ``prestigious'' programs
preferentially populate the same, it is also clear that an abundance of 
opportunities exist at smaller ``non-prestigious'' (and, sometimes, 
non-degree granting) institutions, liberal
arts colleges, government, and industry.  The latter, of course, generally
carry 
enhanced administrative and/or teaching duties, but, on the other hand,
do not entirely preclude a role in the research community.
A Kepler-Meier survival analysis of two disparate institutes demonstrates that
``success'' is a dynamical entity, and that blind consideration of a 20-year
baseline sample can mask important recent trends.
Within ten years of PhD receipt, an equilibrium is reached in which $\sim 45$\%
of the graduates are in identifiably permanent positions, $\sim 20$\% remain in
soft-money positions, and $\sim 35$\% have left research entirely.
Graduates of American universities are $\gtrsim 2\rightarrow 
3\times$ more likely to find
permanent employment in the USA than Canadian or 
Australian graduates are within their respective institute's country.
While the number of American, Canadian, and Dutch PhDs
have grown $\sim 20$\% during the
past decade, the growth in Australia has been closer to $\sim 70$\%.
\end{abstract}



\section{Introduction}
\label{intro}

Attempting to assign any sort of quantitative
ranking scheme, as a measure of an astronomy
graduate program's ``quality'', is a thankless task, and one almost
certainly bound to be met with controversy, if not outright derision.  
Regardless, 
such rankings are, in fact, published on a semi-regular basis, most noticeably
as part of the Gourman (1997) and U.S. News Graduate School Rankings ({\tt
http://www.usnews.com/usnews/edu/bey-\break
ond/}).  In both
cases, though, the rankings are effectively a subjective measure of the
astronomical community's perception of each school's graduate 
program.\footnote{Such subjective biases are, unfortunately, a reality students
should be aware of, although we do not wish to dwell upon them here.}

Very few attempts at objectively quantifying the quality of a given astronomy
graduate program exist, the two most notable being Domen
\& Thronson (1988) and Trimble (1991).   In the former, Domen \& Thronson
conclude that, in the mean, graduates of Berkeley, Harvard, Caltech, Princeton,
and Chicago, were more than six times as successful as graduates of Arizona,
UCLA, Colorado, Minnesota, and Virginia,\footnote{The first versus bottom five
entrants of their Table II.} in obtaining junior\footnote{By ``junior'', we
mean associate and assistant professorships.} faculty positions at the 32 major
institutes comprising their survey sample.  In the latter, Trimble concludes
that while $\sim 80$\% of graduates from one ``prestigious'' university are
typically still involved in astronomy research, the fraction from a comparison
``non-prestigious'' program is only $\sim 50$\%.\footnote{These percentages
are based upon Trimble's (1991) Table 1, combining the entries for graduates
employed at PhD-granting \it and \rm government (or industrial) labs, along
with the small fraction of graduates working in support duties (both hardware
and software) who still maintain a modest publication record.  The resulting
sum should (roughly) parallel our selection criteria, as described in Section
\ref{method}.}  

Both the Domen \& Thronson (1988) and Trimble (1991) studies provide crucial
quantitative
evidence in support of the hypothesis of a hierarchy of quality in astronomy
graduate programs, although our contention is that 
there are some subtle effects within the numbers which 
suggest the situation is perhaps 
not as grim as it would appear on the surface.
For example, Domen \& Thronson only
consider the PhD source for professorial employees of the 32 large
degree-granting institutes in the USA - i.e., the sample is biased, to some
degree.  One is left wondering about those
graduates employed at the numerous small colleges and non-degree-granting
universities, along with those in industry and government laboratories, who
still remain active members of the astronomy research community.  Conversely,
Trimble's comparison is based upon only two universities, and one might
ask if these two are truly representative of the community at large.
Addressing the (potential) shortcomings of these earlier studies is the
focus of our current analysis.

In Section \ref{method}, we describe the methodology employed in determining
the present-day (circa December 1998) status of 897 1975-1994 
astronomy PhD graduates from
19 different schools in 4 different countries.  While clearly not intended
to be 100\% complete\footnote{Although, our sample does represent $\sim 1/4$
of the astronomy PhDs granted during the 20-year 1975-1994 baseline.}, the
sample is fairly representative and unbiased,
including examples of those schools
traditionally considered as prestigious,
as well as lesser-appreciated
large and small programs, from several countries.  
Our putative ranking scheme is a simple objective one, based solely upon 
the fraction of a given school's PhD graduates who are still involved in
astronomical research today, regardless of the perceived status of a given
researcher's present-day institution.  More complicated schemes could be
envisioned whereby additional weight is ascribed to, say, publication
frequency, citation history,
grant application and/or observing proposal success rates, etc, but we are
strongly of the opinion that \it 
the simplest, and perhaps ultimate, measure of a
given program, is the success rate of its graduates in finding long-term
astronomical research careers\rm.

The main results and conclusions of our study are drawn in Section 
\ref{discussion}, and summarized in Section \ref{summary}.

\section{Methodology}
\label{method}

Ideally, one would like to examine the success rates of \it all \rm the
astronomy and astrophysics
degree-granting institutions.\footnote{Doubly so, since it is safe to say that
the first thing every reader of this paper will do is look for their own
institution in Tables 1 and
2!}  Unfortunately, with 4695 theses listed in NASA's
Astrophysics Data System (ADS) database (for
the 1975-1994 baseline of our study), of which 3700 were classified
as ``astronomy/astrophysics'',\footnote{Eliminating the $\sim 20$\% of
ADS-listed PhDs which, for example, are classified as particle physics,
atmospheric, solar, or 
lunar/planetary; we should stress though that our conclusions
are \it not \rm dependent upon the exclusion of these ``non-astronomy'' PhDs
from the sample.},
100\% completeness was just not feasible.
Instead, representative large versus small, and prestigious versus
non-prestigious, institutions were randomly selected for study.  For two of
the universities in our study, the ADS list of PhD graduates was compared
against the respective institutes' official records, and found to be 100\%
complete.

For the American universities, the top three from
the 1998 U.S.
News Astrophysics Graduate School Rankings (Caltech,
Princeton and Harvard), three
from just outside the top ten (Colorado, Maryland and Hawaii), 
and two unranked
(New Mexico State (NMSU) and Wyoming) programs 
were selected.  These eight schools accounted for $\sim 1/7$
of the ADS-listed astronomy PhDs granted during 1975-1994.  For the Canadian
sample, we simply included all eight
universities which granted more than ten PhDs during the same 20-year baseline.
For these 16 North American institutes, their respective 1975-1994 graduate
lists were culled from the ADS master list of 3700.

The non-North American institutions were more problematic, as the vast majority
are either not included in ADS (the norm) or woefully incomplete - as such,
only three universities fall into this category, two from Australia (Mount
Stromlo \& Siding Spring Observatories and Sydney University) and our sole
European entrant (Leiden).  For these three, we either had local access to
a complete set of Annual Reports (within which year-by-year graduate
information was available) or a contact at the institute in question had the
relevant information in a readily available electronic form.
While it
would have been desirable to include, for example, several U.K. universities,
year-by-year graduate information was not available to the authors.

In Table 1, the 19 institutions in our study are listed in descending order of
total number of astronomy PhDs granted ($n_{\rm g}$)
during 1975-1994 (column 14); over an order of magnitude difference exists
between the largest and smallest program.  Five-year sub-samples 
(columns 2,5,8,11) are provided, demonstrating the overall trend of increased
astronomer production, a point to which we return in Section \ref{discussion}.
Again, the 897 graduates tracked in this analysis
represent $\sim 1/4$ of the total number of 
astronomy PhDs (3700) in the ADS; our
sample should be a fairly representative one of the population as a whole.

\small
\begin{table*}
\caption{Number of astronomy and astrophysics 
PhD graduates ($n_{\rm g}$) per institution, along
with the number still involved in astronomical research (n$_{\rm
astr}$), as of late-1998, and the number with identifiably
permanent or tenure-track (n$_{\rm perm}$) positions.  The number of 
graduates of uncertain status are noted in parenthesis.}
\begin{center}
\begin{tabular}{lp{0.16in}p{0.35in}p{0.25in}p{0.16in}p{0.35in}p{0.25in}p{0.16in}p{0.35in}p{0.25in}p{0.16in}p{0.35in}p{0.25in}p{0.16in}p{0.35in}p{0.25in}}\hline
Institution & \multicolumn{3}{c}{\hspace{-0.2in}1975-1979} & 
\multicolumn{3}{c}{\hspace{-0.2in}1980-1984} & 
\multicolumn{3}{c}{\hspace{-0.2in}1985-1989} & 
\multicolumn{3}{c}{\hspace{-0.2in}1990-1994} & 
\multicolumn{3}{c}{\hspace{-0.2in}Total} \\
& n$_{\rm g}$ & \hspace{-0.10in}n$_{\rm astr}$ & \hspace{-0.20in}n$_{\rm perm}$
& n$_{\rm g}$ & \hspace{-0.10in}n$_{\rm astr}$ & \hspace{-0.20in}n$_{\rm perm}$
& n$_{\rm g}$ & \hspace{-0.10in}n$_{\rm astr}$ & \hspace{-0.20in}n$_{\rm perm}$
& n$_{\rm g}$ & \hspace{-0.10in}n$_{\rm astr}$ & \hspace{-0.20in}n$_{\rm perm}$
& n$_{\rm g}$ & \hspace{-0.10in}n$_{\rm astr}$ & \hspace{-0.20in}n$_{\rm perm}$
\\\hline
Caltech     & 
   38 & \hspace{-0.065in}29    &  \hspace{-0.175in}21(7) & 
   31 & \hspace{-0.065in}21    &  \hspace{-0.175in}17(2) & 
   26 & \hspace{-0.065in}19    &  \hspace{-0.175in}11(1) & 
   37 & \hspace{-0.065in}21    &  \hspace{-0.11in}8(1) &
  \hspace{-0.055in}132 & \hspace{-0.065in}90    & \hspace{-0.175in}57(11)\\
Princeton   & 
   25 & \hspace{-0.065in}18    & \hspace{-0.175in}14(2) & 
   20 & \hspace{-0.065in}13    &  \hspace{-0.11in}7(3) & 
   27 & \hspace{-0.065in}22    & \hspace{-0.175in}15(1) & 
   29 & \hspace{-0.065in}21    & \hspace{-0.175in}10(2) &
  \hspace{-0.055in}101 & \hspace{-0.065in}74    & \hspace{-0.175in}46(8) \\
Harvard     & 
   26 & \hspace{-0.065in}19    & \hspace{-0.175in}16(2) & 
   20 & \hspace{-0.065in}17    & \hspace{-0.175in}13(2) & 
   24 & \hspace{-0.065in}17    & \hspace{-0.175in}10(2) & 
   21 & \hspace{-0.065in}17    &  \hspace{-0.11in}5(2) &
                  91  & \hspace{-0.065in}70    & \hspace{-0.175in}44(8)\\
Leiden      & 
   18 & \hspace{-0.065in}11    & \hspace{-0.11in}9(2) & 
   22 & \hspace{-0.065in}14    & \hspace{-0.175in}13    & 
   26 & \hspace{-0.065in}14    & \hspace{-0.175in}10(3) & 
   24 & \hspace{-0.065in}14    &  \hspace{-0.11in}2(2) &
                  90 & \hspace{-0.065in}53    & \hspace{-0.175in}34(7) \\
Maryland    & 
   26 & \hspace{-0.065in}14    &  \hspace{-0.11in}9(4) & 
   16 & \hspace{-0.065in}11    &  \hspace{-0.11in}5(4) & 
   22 & \hspace{-0.065in}17    &  \hspace{-0.11in}5(6) & 
   20 & \hspace{-0.065in}18    &  \hspace{-0.11in}5(1) &
                  84 & \hspace{-0.065in}60    & \hspace{-0.175in}24(15)\\
MSSSO       & 
\hspace{0.062in}9 &  \hspace{-0.005in}6    &  \hspace{-0.11in}5    & 
   18 & \hspace{-0.065in}11    &  \hspace{-0.11in}6    & 
   16 &  \hspace{-0.005in}9    &  \hspace{-0.11in}6    & 
   25 & \hspace{-0.065in}15    &  \hspace{-0.11in}3    &
                 68 & \hspace{-0.065in}41    & \hspace{-0.175in}20    \\
Colorado    & 
   12 & \hspace{-0.005in}5     & \hspace{-0.11in}5     & 
   10 & \hspace{-0.005in}8     &  \hspace{-0.11in}6(1) & 
   24 & \hspace{-0.065in}16(2) & \hspace{-0.11in}8(3)  & 
   20 & \hspace{-0.065in}14(2) &  \hspace{-0.11in}3(1) &
                  66  & \hspace{-0.065in}43(4) & \hspace{-0.175in}22(5)\\
Toronto     & 
   15 &  \hspace{-0.005in}7    &  \hspace{-0.11in}6    & 
   18 &  \hspace{-0.005in}8    &  \hspace{-0.11in}6    & 
   17 &  \hspace{-0.005in}9    &  \hspace{-0.11in}8    & 
   11 &  \hspace{-0.005in}6    &  \hspace{-0.11in}1(1) &
                  61 & \hspace{-0.065in}30    & \hspace{-0.175in}21(1) \\
Sydney      & 
   12 &  \hspace{-0.005in}4    &  \hspace{-0.11in}4    & 
    \hspace{0.062in}3 &  \hspace{-0.005in}1    &  \hspace{-0.11in}1    & 
   14 & \hspace{-0.065in}12    &  \hspace{-0.11in}5    & 
   17 &  \hspace{-0.005in}9(1) &  \hspace{-0.11in}2    &
                  46 & \hspace{-0.065in}26(1) & \hspace{-0.175in}12    \\
Hawaii      & 
    \hspace{0.062in}6 &  \hspace{-0.005in}5    &  \hspace{-0.11in}4(1) & 
    \hspace{0.062in}3 &  \hspace{-0.005in}2    &  \hspace{-0.11in}1    & 
    \hspace{0.062in}3 &  \hspace{-0.005in}3    &  \hspace{-0.11in}2(1) & 
                   14 &  \hspace{-0.065in}10    &  \hspace{-0.11in}1    &
                   26 &  \hspace{-0.065in}20   &  \hspace{-0.113in}8(2)  \\
NMSU        & 
    \hspace{0.062in}6 &  \hspace{-0.005in}3(1) &  \hspace{-0.11in}2(2) & 
    \hspace{0.062in}2 &  \hspace{-0.005in}0(1) &  \hspace{-0.11in}0(1) & 
    \hspace{0.062in}9 &  \hspace{-0.005in}6    &  \hspace{-0.11in}3(1) & 
    \hspace{0.062in}5 &  \hspace{-0.005in}3    &  \hspace{-0.11in}0(1) &
                   22 &  \hspace{-0.065in}12(2)&  \hspace{-0.113in}5(5) \\
UBC         & 
    \hspace{0.062in}8 &  \hspace{-0.005in}5    &  \hspace{-0.11in}4(1) & 
    \hspace{0.062in}3 &  \hspace{-0.005in}1    &  \hspace{-0.11in}1    & 
    \hspace{0.062in}3 &  \hspace{-0.005in}3    &  \hspace{-0.11in}1    & 
    \hspace{0.062in}7 &  \hspace{-0.005in}4    &  \hspace{-0.11in}1    &
                   21 &  \hspace{-0.065in}13   &  \hspace{-0.113in}7(1) \\
Western     & 
    \hspace{0.062in}7 &  \hspace{-0.005in}2    &  \hspace{-0.11in}2    & 
    \hspace{0.062in}2 &  \hspace{-0.005in}0    &  \hspace{-0.11in}0    & 
    \hspace{0.062in}4 &  \hspace{-0.005in}3    &  \hspace{-0.11in}1(1) & 
    \hspace{0.062in}4 &  \hspace{-0.005in}3    &  \hspace{-0.11in}1    &
                   17 &  \hspace{-0.005in}8    &  \hspace{-0.113in}4(1) \\
Montreal    & 
    \hspace{0.062in}0 &  \hspace{-0.005in}0    &  \hspace{-0.11in}0    & 
    \hspace{0.062in}1 &  \hspace{-0.005in}0    &  \hspace{-0.11in}0    & 
    \hspace{0.062in}2 &  \hspace{-0.005in}2    &  \hspace{-0.11in}0(1) & 
   11 &  \hspace{-0.005in}9    &  \hspace{-0.11in}1(2) &
                   14 &  \hspace{-0.065in}11   &  \hspace{-0.113in}1(3) \\
Wyoming     & 
    \hspace{0.062in}1 &  \hspace{-0.005in}0    &  \hspace{-0.11in}0    & 
    \hspace{0.062in}5 &  \hspace{-0.005in}3    &  \hspace{-0.11in}2(1) & 
    \hspace{0.062in}1 &  \hspace{-0.005in}1    &  \hspace{-0.11in}0    & 
    \hspace{0.062in}7 &  \hspace{-0.005in}5    &  \hspace{-0.11in}1(1) &
                   14 &  \hspace{-0.005in}9    &  \hspace{-0.113in}3(2) \\
Victoria    & 
    \hspace{0.062in}1 &  \hspace{-0.005in}1    &  \hspace{-0.11in}1    & 
    \hspace{0.062in}4 &  \hspace{-0.005in}4    &  \hspace{-0.11in}4    & 
    \hspace{0.062in}2 &  \hspace{-0.005in}2    &  \hspace{-0.11in}0(1) & 
    \hspace{0.062in}5 &  \hspace{-0.005in}3    &  \hspace{-0.11in}3    &
                   12 &  \hspace{-0.065in}10   &  \hspace{-0.113in}8(1) \\
Queen's     & 
    \hspace{0.062in}2 &  \hspace{-0.005in}1    &  \hspace{-0.11in}1    & 
    \hspace{0.062in}1 &  \hspace{-0.005in}1    &  \hspace{-0.11in}1    & 
    \hspace{0.062in}3 &  \hspace{-0.005in}1    &  \hspace{-0.11in}1    & 
    \hspace{0.062in}5 &  \hspace{-0.005in}2    &  \hspace{-0.11in}0    &
                   11 &  \hspace{-0.005in}5    &  \hspace{-0.113in}3    \\
York        &  
    \hspace{0.062in}2 &  \hspace{-0.005in}0(1) &  \hspace{-0.11in}0    &  
    \hspace{0.062in}3 &  \hspace{-0.005in}0    &  \hspace{-0.11in}0    &  
    \hspace{0.062in}1 &  \hspace{-0.005in}0    &  \hspace{-0.11in}0    & 
    \hspace{0.062in}5 &  \hspace{-0.005in}2    &  \hspace{-0.11in}0    &
                   11 &  \hspace{-0.005in}2(1) &  \hspace{-0.113in}0    \\
Calgary     &  
    \hspace{0.062in}2 &  \hspace{-0.005in}1    &  \hspace{-0.11in}1    &  
    \hspace{0.062in}2 &  \hspace{-0.005in}1    &  \hspace{-0.11in}1    &  
    \hspace{0.062in}1 &  \hspace{-0.005in}1    &  \hspace{-0.11in}0(1) & 
    \hspace{0.062in}5 &  \hspace{-0.005in}3    &  \hspace{-0.11in}1    &
                   10 &  \hspace{-0.005in}6    &  \hspace{-0.113in}3(1) \\\hline
Total       &  
\hspace{-0.060in}216 & \hspace{-0.135in}131(2) & \hspace{-0.23in}104(21) &  
\hspace{-0.065in}184 & \hspace{-0.135in}116(1) & \hspace{-0.165in}84(14)  &  
\hspace{-0.060in}225 & \hspace{-0.135in}157(2) & \hspace{-0.165in}86(22)  & 
\hspace{-0.060in}272 & \hspace{-0.132in}179(3) & \hspace{-0.165in}48(14)  &
\hspace{-0.065in}897 & \hspace{-0.132in}583(8) & \hspace{-0.233in}322(71) \\\hline
\end{tabular}
\end{center}
\end{table*}
\normalsize

The next, and most arduous, step in ultimately determining a given program's
success rate, was determining the whereabouts and/or astronomy ``status''
of the 897 PhD recipients listed in column 14 of Table 1.  The methodology
employed was typically as follows:  (i) ADS, the Science Citation Index
(SCI), or the Institute for Scientific Information Citation Database 
(ISI)\footnote{The SCI and ISI are clearly superior for tracking 
down astronomy graduates whose publication habits, for one reason or another,
avoid the traditional journals monitored by the ADS.}
were searched for the most
recent publication(s) - no preference was made as to an author's position in 
the author list; (ii) institutional affiliation and email address were
noted; (iii) confirmation of institutional affiliation was ensured in all
cases, either by relevant web page or email contact; (iv) institutional status
(i.e., soft-money, fixed-term contract, tenured faculty, open-ended civil
appointment, etc.) was confirmed in \it all \rm
cases, again, either by web page or
email contact.  In total, $\sim 1/3$ of our total sample were contacted via
email, the majority of whom provided clarification where necessary. 
No subjective classification based upon publication frequency was 
incorporated into the analysis; provided a present-day (late-1998)
institutional affiliation could be confirmed, a graduate included in a
given $n_{\rm g}$ entry in Table 1 would then also be counted in
the corresponding $n_{\rm astr}$ entry.

Occasionally, an unequivocal conclusion regarding a given graduate's status in
the community was impossible; those falling into this category\footnote{Usually
those who had either
published a paper during 1997-1998, but for whom no confirmation
of current institutional affiliation could be made, or those few who did not
respond to the email request for institutional and/or 
job status clarification.} are 
included by their relevant entry of Table 1 in parantheses.

Column 15 of Table 1 lists the number of 1975-1994 graduates still involved in
astronomical research ($n_{\rm astr}$), for each of the institutions, while
column 16 provides the number which are currently (as of late-1998) in
identifiably permanent astronomy research positions.  The latter are 
overwhelmingly based at university or national facilities, but occasionally may
include industry positions if they are clearly related to astronomy 
research\footnote{Such positions are generally astronomical software
or instrumentation development-related, 
the research for which is published, and not
restricted to ``in-house'' documents.}.  Again, five-year sub-totals are also
provided under the relevant columns.  

\section{Discussion}
\label{discussion}

Before discussing any putative ranking based upon the production of long-term
career astronomers, several general trends from Table 1 should be noted.

First, the five-year sub-samples show
that the number of 1985-1994 graduates (columns 8 and 11) 
is $\sim 25$\% greater than the
number of 1975-1984 (columns 2 and 5) graduates, growth which, in North
America, is reflected
primarily in what were once the smaller North American programs (e.g., 
Hawaii, NMSU, Montreal, and Quuen's), a trend already noted by 
Domen \& Thronson (1988).

\small
\begin{table*}
\caption{Fraction of PhD graduates still
involved in astronomical research (f$_{\rm
astr}$), as of late-1998, along with the fraction of graduates categorised as 
permanent or tenure-track (f$_{\rm perm}$).}
\begin{center}
\begin{tabular}{lp{0.45in}p{0.52in}p{0.45in}p{0.52in}p{0.45in}p{0.52in}p{0.45in}p{0.52in}p{0.45in}p{0.32in}}\hline
Institution & \multicolumn{2}{c}{\hspace{-0.2in}1975-1979} & 
\multicolumn{2}{c}{\hspace{-0.2in}1980-1984} & 
\multicolumn{2}{c}{\hspace{-0.2in}1985-1989} & 
\multicolumn{2}{c}{\hspace{-0.2in}1990-1994} & 
\multicolumn{2}{c}{\hspace{-0.1in}Total} \\
& f$_{\rm astr}$ & \hspace{-0.05in}f$_{\rm perm}$
& f$_{\rm astr}$ & \hspace{-0.05in}f$_{\rm perm}$
& f$_{\rm astr}$ & \hspace{-0.05in}f$_{\rm perm}$
& f$_{\rm astr}$ & \hspace{-0.05in}f$_{\rm perm}$
& f$_{\rm astr}$ & \hspace{-0.05in}f$_{\rm perm}$ \\\hline
Victoria               & 
\hspace{-0.08in}1.00   & \hspace{-0.18in}1.00          &
\hspace{-0.08in}1.00   & \hspace{-0.18in}1.00          &
\hspace{-0.08in}1.00   & \hspace{-0.1in}.00-.50        &
        .60            & \hspace{-0.1in}.60            &
   \underline{.83}     & \hspace{-0.1in}\underline{.67}-.75 \\
Montreal               & 
        .00            & \hspace{-0.1in}.00            &
        .00            & \hspace{-0.1in}.00            &
\hspace{-0.08in}1.00   & \hspace{-0.1in}.00-.50        &
        .82            & \hspace{-0.1in}.09-.27        &
   \underline{.79}     & \hspace{-0.1in}\underline{.07}-.29 \\
Harvard                & 
        .73            & \hspace{-0.1in}.62-.69        &
        .85            & \hspace{-0.1in}.65-.75        &
        .71            & \hspace{-0.1in}.42-.50        &
        .81            & \hspace{-0.1in}.24-.33        &
   \underline{.77}     & \hspace{-0.1in}\underline{.48}-.57 \\
Hawaii                 & 
        .83            & \hspace{-0.1in}.67-.83        &
        .67            & \hspace{-0.1in}.33            &
\hspace{-0.08in}1.00   & \hspace{-0.1in}.67-1.0        &
        .71            & \hspace{-0.1in}.07            &
   \underline{.77}     & \hspace{-0.1in}\underline{.31}-.38 \\
Princeton              & 
        .72            & \hspace{-0.1in}.56-.64        &
        .65            & \hspace{-0.1in}.35-.50        &
        .81            & \hspace{-0.1in}.56-.59        &
        .72            & \hspace{-0.1in}.34-.41        &
   \underline{.73}     & \hspace{-0.1in}\underline{.46}-.53 \\
Maryland               & 
        .54            & \hspace{-0.1in}.35-.50        &
        .69            & \hspace{-0.1in}.31-.56        &
        .77            & \hspace{-0.1in}.23-.50        &
        .90            & \hspace{-0.1in}.25-.30        &
   \underline{.71}     & \hspace{-0.1in}\underline{.29}-.46 \\
Caltech                & 
        .76            & \hspace{-0.1in}.55-.74        &
        .67            & \hspace{-0.1in}.55-.61        &
        .73            & \hspace{-0.1in}.42-.46        &
        .57            & \hspace{-0.1in}.22-.24        &
   \underline{.68}     & \hspace{-0.1in}\underline{.43}-.52 \\
Colorado               & 
        .42            & \hspace{-0.1in}.42            &
        .80            & \hspace{-0.1in}.60-.70        &
        .67-.75        & \hspace{-0.1in}.33-.46        &
        .70-.80        & \hspace{-0.1in}.15-.20        &
   \underline{.65}-.71 & \hspace{-0.1in}\underline{.33}-.41 \\
Wyoming                & 
        .00            & \hspace{-0.1in}.00            &
        .60            & \hspace{-0.1in}.40-.60        &
\hspace{-0.08in}1.00   & \hspace{-0.1in}.00            &
        .71            & \hspace{-0.1in}.14-.29        &
   \underline{.64}     & \hspace{-0.1in}\underline{.21}-.36 \\
UBC                    & 
        .63            & \hspace{-0.1in}.50-.63        &
        .33            & \hspace{-0.1in}.33            &
\hspace{-0.08in}1.00   & \hspace{-0.1in}.33            &
        .57            & \hspace{-0.1in}.14            &
   \underline{.62}     & \hspace{-0.1in}\underline{.33}-.38 \\
MSSSO                  & 
        .67            & \hspace{-0.1in}.56            &
        .61            & \hspace{-0.1in}.33            &
        .56            & \hspace{-0.1in}.38            &
        .60            & \hspace{-0.1in}.12            &
   \underline{.60}     & \hspace{-0.1in}\underline{.29}     \\
Calgary                & 
        .50            & \hspace{-0.1in}.50            &
        .50            & \hspace{-0.1in}.50            &
\hspace{-0.08in}1.00   & \hspace{-0.1in}.00-1.0       &
        .60            & \hspace{-0.1in}.20            &
   \underline{.60}     & \hspace{-0.1in}\underline{.30}-.40 \\
Leiden                 & 
        .61            & \hspace{-0.1in}.50-.61        &
        .64            & \hspace{-0.1in}.59            &
        .54            & \hspace{-0.1in}.38-.50        &
        .58            & \hspace{-0.1in}.08-.17        &
   \underline{.59}     & \hspace{-0.1in}\underline{.38}-.46 \\
Sydney                 & 
        .33            & \hspace{-0.1in}.33            &
        .33            & \hspace{-0.1in}.33            &
        .86            & \hspace{-0.1in}.36            &
        .53-.59        & \hspace{-0.1in}.12            &
   \underline{.57}-.59 & \hspace{-0.1in}\underline{.26}     \\
NMSU                   & 
        .50-.67        & \hspace{-0.1in}.33-.67        &
        .00-.50        & \hspace{-0.1in}.00-.50        &
        .67            & \hspace{-0.1in}.33-.44        &
        .60            & \hspace{-0.1in}.00-.20        &
   \underline{.55}-.64 & \hspace{-0.1in}\underline{.23}-.45 \\
Toronto                & 
        .47            & \hspace{-0.1in}.40            &
        .44            & \hspace{-0.1in}.33            &
        .53            & \hspace{-0.1in}.47            &
        .55            & \hspace{-0.1in}.09-.18        &
   \underline{.49}     & \hspace{-0.1in}\underline{.34}-.36 \\
Western                & 
        .29            & \hspace{-0.1in}.29            &
        .00            & \hspace{-0.1in}.00            &
        .75            & \hspace{-0.1in}.25-.50        &
        .75            & \hspace{-0.1in}.25            &
   \underline{.47}     & \hspace{-0.1in}\underline{.24}-.29 \\
Queen's                & 
        .50            & \hspace{-0.1in}.50            &
\hspace{-0.08in}1.00            & \hspace{-0.18in}1.00            &
        .33            & \hspace{-0.1in}.33            &
        .40            & \hspace{-0.1in}.00            &
   \underline{.45}     & \hspace{-0.1in}\underline{.27}     \\
York                   & 
        .00-.50        & \hspace{-0.1in}.00            &
        .00            & \hspace{-0.1in}.00            &
        .00            & \hspace{-0.1in}.00            &
        .40            & \hspace{-0.1in}.00            &
   \underline{.18}-.27 & \hspace{-0.1in}\underline{.00}     \\\hline
Total                  & 
        .61-.62        & \hspace{-0.1in}.48-.58        &
        .63-.64        & \hspace{-0.1in}.46-.53        &
        .70-.71        & \hspace{-0.1in}.38-.48        &
        .66-.67        & \hspace{-0.1in}.18-.23        &
   \underline{.65}-.66 & \hspace{-0.1in}\underline{.36}-.44 \\\hline
\end{tabular}
\end{center}
\end{table*}
\normalsize

Second, and of particular interest (concern?)
for Australian astronomy, is the fact that MSSSO and Sydney have increased
their number of PhD graduates by $\sim 50$\% and $\sim 100$\%, respectively,
from 1975-1984 to 1985-1994.
These two institutions dominate astronomy PhD production in Australia,
accounting for 50\% of the total
(Table 3.5.2 of ``Australian Astronomy: Beyond
2000''\footnote{After removal of the Monash University entry, which reflects
the pure mathematics department as a whole, and not the small subset of
astronomers therein.}).
Of concern should be the fact that this $\sim 70$\% increase in
the number of PhDs granted has not been matched by a parallel increase in
the number of postdoctoral and permanent faculty/research positions.  For
comparison, during the same time period, the number of American,
Canadian, and Dutch PhDs increased by the more modest (but not 
insubstantial) $\sim 20$\%.

Shear PhD production (column 14 of Table 1), while (perhaps)
a useful measure at some level, should not be the only yardstick by which a
given program's career
astronomer-production ``efficiency'' is measured.  More useful
is normalizing the present-day number of active (both soft-money and permanent)
research astronomers produced
by each institution ($n_{\rm astr}$ and $n_{\rm perm}$
of Table 1), by the number
of PhDs awarded ($n_{\rm g}$).

In Table 2, this fraction of total astronomy PhD graduates still involved in
astronomy research f$_{\rm astr}$, along with the fraction possessing an
identifiably permanent or tenured position f$_{\rm perm}$, are listed for each
of the 19 institutions in our study.  Columns 10 and 11 give the numbers based
upon the entire 1975-1994 baseline, while the five-year sub-sample fractions
are likewise tabulated in the appropriate columns.  The parenthesized
uncertainties of Table 1 are reflected by hyphenated ranges in Table 2 (i.e., 
the underlined lower limits of columns 10 and 11 are certain), but some leeway
(particularly as far as f$_{\rm perm}$ goes) remains, due to unresolved
permanence-vs-non-permanence issues, for some graduates.  Sorting of the
institutions in Table 2 was done, in descending order, by present-day fraction
of astronomers still active in the field (i.e., column 10).

Many caveats must be borne in mind when interpreting Table 2.
First, and perhaps foremost, we must caution the reader against
over-interpreting the rankings of the six small Canadian universities (the
final six entrants of Table 1).  With $<1$ PhD granted per year, during this
20-year period, their rankings are subject to the whims of small number
statistics.  Second, no provision (of course) is made for those who willingly
left the field to pursue non-astronomy career goals; indeed, considering the
present-day astronomy job market, and the PhD overproduction rate (Thronson
1991), programs which responsibly inform graduate students about the realities
of the market, and offer parallel non-astronomy skills training, could end up
suffering in the rankings in Table 2.\footnote{On the other hand, so very few
institutions, in our experience, present such options, that the cynic might say
that such a putative effect is entirely non-existent!}  Third, it became
readily apparent that certain schools have significant foreign student
enrollment; due to the nature of \it some \rm overseas fellowships, this
occasionally allows young PhD recipients to return to their home country to an
early faculty/permanent position.  Again, some schools have
their f$_{\rm perm}$ enhanced by this mechanism, but in general the effect is
small.  A (perhaps) more impartial measure of f$_{\rm perm}$ would be to
compare each institute's graduate fraction who find permanent employment in
that institute's country; we will return to this point at the end of this
section.

Caveats of a more scientific nature are equally important to recall.  
First, averaging over any
given institute's output can be potentially misleading;  each school has their
sphere of expertise, and at some level, one must worry about comparing an
f$_{\rm astr}$ from an instrumentation-dominated program, with an f$_{\rm astr}$
from a theoretical cosmology program.  Just as important, within any given
school, there will exist a hierarchy in the success rates of particular PhD
supervisors in training long-term career astronomers.  This averaging over
areas of expertise (and non-expertise) and individual faculty supervisors will
not do justice to that lone supervisor who continually trains successful
graduates, but who toils in relative obscurity in an otherwise mediocre
department.  Prospective students, for example, should bear these latter points
in mind when investigating graduate school options, and should
not blindly adhere to
rankings of the like presented in Table 2, the annual U.S. News Gradute
School or Gourman Rankings.

Having presented the numerous caveats, though, there is still much to be
learned from Table 2.  Perhaps the most (pleasantly) surprising result 
of our analysis is that, with very few exceptions, \it
$\sim 55\rightarrow 75$\% of all the graduates, regardless of
original
institutional affiliation, are still involved in astronomical research. \rm
In total, 583 of the 897 PhD graduates in our study, or $\sim 65$\%, were
still involved in research, \it at some level\rm, in late-1998.

Examining the five-year sub-samples, we see that there is
little variation in f$_{\rm astr}$, at the present-day, regardless of
graduation date.  The percentages for 1975-1979, 1980-1984, 1985-1989, and
1990-1994, are 61\%, 63\%, 70\%, and 66\%, respectively; we initially assumed
the constancy in these f$_{\rm astr}$ values suggested that the
decision to pursue non-astronomical interests was made within a few years of 
PhD receipt, with only marginal migration from the field
thereafter.  While this is perhaps true in some cases, we are now of the
opinion that the similarity is more a conspiracy in the temporal evolution
of the ``drop-out rate'', in that research ``half-life'' for graduates from the
1970s was longer than those from the 1990s, the result of which is the f$_{\rm
astr}\approx 0.65$, for each of the five-year sub-samples of Table 2.  

The first three of the five-year
sub-samples of Table 2
also show similar fractions of astronomers now in permanent
positions f$_{\rm perm}$, 
typically $\sim 40\rightarrow 50$\%; apparently, within ten years of
graduation, an equilibrium has been reached in which the ratio of astronomy
graduates with permanent positions to graduates on soft-money to graduates who
have left research entirely\footnote{Modulo our definition of ``leaving
research'' described in Section \ref{method}.} is approximately 45:20:35.  
The most recent sub-sample
(1990-1994), as witnessed by the final entry to Table 2, has clearly not
reached this equilibrium, which is not surprising, since many of the graduates
in question would still be in the midst of their second postdoctoral position.

In passing, we note that the Canadian numbers are slightly lower than the
aforementioned ${\rm f}_{\rm astr}=0.65$; 
$\sim 54$\% of Canadian graduates are still
involved in astronomical research ($\sim 30$\% in permanent positions).  
Canada's marginally poorer performance, in this regard,
should not be overly surprising, when one takes into account the dismal funding
situation faced by Canadian astronomers (van der Kruit 1994).

Regarding the implications for Australian astronomy - 
first, both the overall
fraction still in astronomy ($\sim 59$\%) and the fraction with permanent
positions ($\sim 28$\%), are only marginally lower than those for the entire
sample ($\sim 65$\% and $\sim 36$\%, respectively).
What is perhaps of some concern is that the fraction of graduates who have
eventually found permanent positions \it within \rm Australia, as derived from
Table 3, is only $\sim
11$\%; $\sim 20$\% of the 1975-1994
Sydney graduates currently fall into this category, while the fraction for
MSSSO graduates is $\sim 6$\%.  While this 6\% rate is one of
the lowest of the
19 institutions in this study (Table 3), 
MSSSO compensates by training a very high
fraction of graduates who eventually fill permanent overseas
positions ($\sim 24$\%).
In comparison, $\sim 17$\% of Canadian university astronomy
graduates have settled into permanent positions within Canada, $\sim 21$\% of
Leiden graduates now have permanent positions within the Netherlands, and $\sim
34$\% of American institutional graduates now have permanent positions with the
USA.

\small
\begin{table}[ht]
\caption{Fraction of PhD graduates with permanent, or tenure-track, astronomy
research positions within the same country as their PhD institution ($f_{\rm
perm}^{\rm same}$) - derived from number of graduates in this category
($n_{\rm perm}^{\rm same}$) and the total number of graduates ($n_{\rm g}$).}
\begin{center}
\begin{tabular}{lrrr}\hline
Institution & $n_{\rm g}$ & $n_{\rm perm}^{\rm same}$ & $f_{\rm perm}^{\rm
same}$ \\\hline
Victoria    &   12 &    7$\;\;$ &  .58$\;\;$ \\
Harvard     &   91 &   35$\;\;$ &  .38$\;\;$ \\
Princeton   &  101 &   38$\;\;$ &  .38$\;\;$ \\
Caltech     &  132 &   49$\;\;$ &  .37$\;\;$ \\
Hawaii      &   26 &    8$\;\;$ &  .31$\;\;$ \\
Colorado    &   66 &   20$\;\;$ &  .30$\;\;$ \\
Maryland    &   84 &   22$\;\;$ &  .26$\;\;$ \\
NMSU        &   22 &    5$\;\;$ &  .23$\;\;$ \\
Wyoming     &   14 &    3$\;\;$ &  .21$\;\;$ \\
Leiden      &   90 &   19$\;\;$ &  .21$\;\;$ \\
Calgary     &   10 &    2$\;\;$ &  .20$\;\;$ \\
Sydney      &   46 &    9$\;\;$ &  .20$\;\;$ \\
UBC         &   21 &    4$\;\;$ &  .19$\;\;$ \\
Queen's     &   11 &    2$\;\;$ &  .18$\;\;$ \\
Western     &   17 &    3$\;\;$ &  .18$\;\;$ \\
Toronto     &   61 &    7$\;\;$ &  .11$\;\;$ \\
Montreal    &   14 &    1$\;\;$ &  .07$\;\;$ \\
MSSSO       &   68 &    4$\;\;$ &  .06$\;\;$ \\
York        &   11 &    0$\;\;$ &  .00$\;\;$ \\\hline
Total       &  897 &  238$\;\;$ &  .27$\;\;$ \\\hline
\end{tabular}
\end{center}
\end{table}
\normalsize

Another interesting aspect of the results of Tables 1-3, for Australian
astronomy, relates back to one
of the conclusions of the ``Australian Astronomy: Beyond 2000'' document
(Section 2.2).  A claim was made therein 
that $\sim 21$\% of Australian PhD graduates obtain
permanent astronomy positions within Australia.\footnote{An estimate was made
that 112 out of 160 Australian astronomy positions are permanent; with an
assumed lifetime per position of 35 years, a claim for 3.2 permanent positions
per year is made.  With (typically) 15 PhDs granted per year, this leads to
their
inferred claim that 21\% of Australian graduates obtain permanent positions
\it within \rm Australia.}
Our analysis of the MSSSO and
Sydney graduates, which, recall, account for half of \it all \rm
Australian astronomy
PhDs, shows that this estimate was a factor of two too
optimistic.  In retrospect this should not be too surprising, since an
underlying assumption of the ``Australian Astronomy: Beyond 2000'' claim was
that all permanent astronomy
positions in Australia are filled by Australian graduates; anecdotally,
this is clearly a flawed assumption, and one which we have now confirmed
quantitatively.

How do our results compare with the earlier Domen \& Thronson (1988)
and Trimble (1991) analyses?  The former clearly demonstrated that the
subset of PhD graduates from the top five programs in the
U.S. News Gradute School Rankings
(Caltech, Princeton, Harvard, Berkeley, and Chicago), in comparison with five
more moderately-ranked programs (Arizona, Colorado, UCLA, Minnesota, and
Virginia), were disproportionately more successful (by more than a factor of
six) in obtaining permanent research positions at the 32 largest PhD
degree-granting institutions.  Our survey was not designed to substantiate or
repudiate Domen \& Thronson, but even a cursory analysis of our
dataset would tend to lend credence to their claim, as would anecdotal wisdom -
this is a negative aspect of such analyses.  On the other hand, we choose to
focus on the positive - our results clearly indicate that $\sim 65$\%$\pm 10$\%
of graduates of \it any \rm astronomy PhD program can expect to maintain a
career in research astronomy - i.e., actual \it source \rm of PhD is of little
importance.\footnote{It is interesting to note that of the seven American
universities ranked below Caltech (the top-ranked program according to the U.S.
News Graduate School Rankings), 
the graduates of four of these (including two from outside the top ten
in the U.S. News rankings) programs actually ranked (marginally) 
higher when it came to maintaining a long-term astronomy research career.}
While it \it is \rm
true, as Domen \& Thronson found, that the graduates
of the so-called prestigious schools may preferentially fill positions at these
same prestigious school, this does \it not \rm preclude the opportunity of
pursuing a research career, quite often, for example,
at one of the \it many \rm non-degree granting colleges and
universities.  While such positions generally carry a heavier
burden of teaching and administrative duties, reduced publishing frequency,
and lead some researchers to feel somewhat isolated from the 
community, they do still allow the determined astronomer to pursue a
research career, albeit (perhaps) at a reduced efficiency from those in the
large, active, degree-granting programs.  This result is perhaps not fully
appreciated upon initial reading of Domen \& Thronson, but is one in which the
prospective graduating student should take heart!

Our conclusions are in mild disagreement with those of Trimble (1991), although
an \it a posteriori \rm examination of her dataset shows that the disagreement
is not as striking as first thought.  Recall first though, that Trimble found
that 18 years after PhD receipt, graduates of one ``prestigious, top four''
(called `P')
university were $\sim 60\rightarrow 100$\% more likely to be active in
astronomy research than the graduates of one ``non-prestigious, second ten''
school (called `NP').  
Our analysis though shows that for the eight US programs in our study,
even the extrema (i.e., Harvard and NMSU) only differ 40\%; for the five large
US programs (i.e., $>2$ PhD recipients per year), Harvard and Colorado differ
by only 15\% (recall Table 2).

In an attempt to uncover the source of the above discrepancy, we returned to
Trimble's (1991) original Kepler-Meier survival curve (her Figure 2) showing
the fraction of graduates from the two institutes in her study still
publishing f$_{\rm pub}$, 
as a function of time since PhD receipt t$_{\rm PhD}$.  While it is
true that her dataset shows that 
graduates of `P' are a factor of two more likely to be publishing at
t$_{\rm PhD}=18$\,yr, than are graduates of `NP', we feel this is an overly
pessimistic reading of the data.  The optimistic interpretation is that
at t$_{\rm PhD}=16$\,yr the ``advantage'' 
enjoyed by graduates of `P' over those of `NP' is only $\sim 30$\%, consistent
with our conclusions.  At t$_{\rm PhD}=17$\,yr, the `NP' curve shows a
precipitous drop in f$_{\rm pub}$, the significance of which should be tempered
with the realization that limited numbers in the yearly bins have now become
important.  While Trimble stressed this t$_{\rm PhD}\ge 17$\,yr discrepancy in
her conclusions, we would counter by stressing the similarity of the samples
for t$_{\rm PhD}<17$\,yr.

We are in a position to perform a similar Kaplan-Meier survival analysis for
our sample, but now covering the 1975-1994 baseline.  To parallel Trimble's
(1991) study, let us contrast the behavior of the traditionally
top-ranked program (Caltech), with that of one chosen from outside the top ten
(Maryland).  The fraction still publishing f$_{\rm pub}$, as a function of time
since PhD receipt t$_{\rm PhD}$, for both programs, is represented in Figure 1
by the filled circles.


\begin{figure}[ht]
\centering
\vspace{3.3in}
\includegraphics{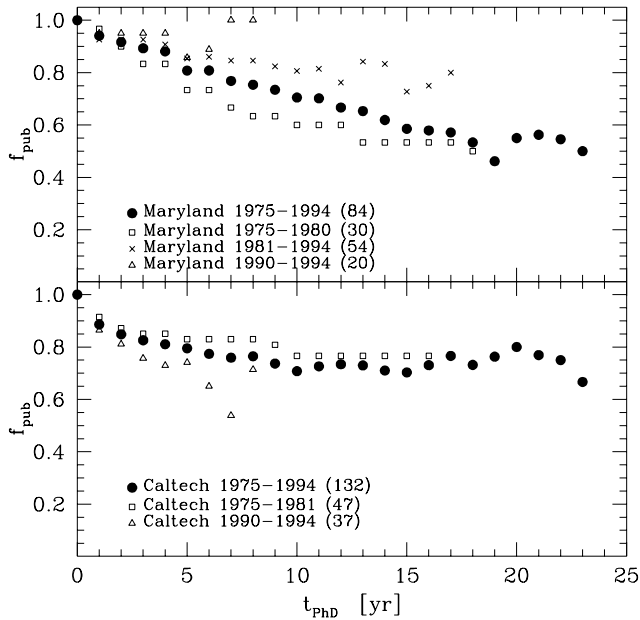}
\caption{Kaplan-Meier survival curves illustrating the fraction of PhD
astronomers f$_{\rm pub}$ still publishing papers, as a function of time since
PhD receipt t$_{\rm PhD}$.  After Trimble (1991), a ``top four'' 
(Caltech) and ``second ten'' (Maryland) program are highlighted.  Even after 20
years, using the full samples of Table 1 (filled circles),
there is only a 30\% difference between the programs; more importantly,
though, adopting the full sample has a tendency to mask recent trends.
Since 1981, the fraction of Maryland graduates still involved in research is
indistinguishable from the Caltech fraction; indeed, during the 1990s, one
might argue that the roles have become reversed.
\label{fig:fig1}}
\end{figure}

The first thing to note from Figure 1 is that we see no precipitous drop in
f$_{\rm pub}$ beyond t$_{\rm PhD}=16$\,yr; over the full baseline, the
difference in f$_{\rm pub}$ between the two programs is typically no more than
30\%.  More importantly, Kaplan-Meier curves 
can mask (not surprisingly) trends occurring on
timescales less than the baseline.  This is particularly evident when
contrasting the behavior of f$_{\rm pub}$, in the Maryland sample, since 1980;
as the crosses in the upper panel demonstrate, this behavior is now
indistinguishable from the Caltech sample, in the lower panel.  In fact, since
1990, one might argue that Maryland graduates have been more successful in
remaining in astronomy, although the numbers per bin
are starting to get small, so we
hesitate to overly interpret this result.  A minor point to note is that $\sim
5\rightarrow 10$\% of graduates immediately leave the research field, never
publishing anything beyond their PhD dissertation.\footnote{Unlike in Trimble
(1991), we have counted the graduate's PhD dissertation as a publication, which
is why that f$_{\rm pub}\equiv 1.0$ at t$_{\rm PhD}\equiv 0$\,yr.}

While we have only shown two representative programs here (Caltech and
Maryland), our ultimate conclusions do not depend on this choice, as was
already evident by the institute-insensitive behavior of f$_{\rm astr}$ seen in
Table 2.

\section{Summary}
\label{summary}

Upon carefully analyzing the present-day status of 897 astronomy PhD
recipients from 19 different institutions, our main conclusions can be
summarized thusly:

\begin{enumerate}
\item Institutional source of one's PhD plays little part in whether or not
(statistically) one will successfully pursue a long-term career in
astronomy - $\sim 55\rightarrow 75$\% of all graduates, regardless of PhD
source, do so.  
\item Graduates of traditionally ``prestigious'' programs are almost certainly 
disproportionately successful in obtaining permanent positions at similarly
ranked schools (echoing Domen \& Thronson 1988), but this has obviously
not precluded
graduates of ``non-prestigious'' programs from pursuing their research careers
from outside the elite, ranked, universities.  Perhaps their efficiency has
been hindered, but they have not had to completely forgo research either.
\item Our results complement those of Domen \& Thronson (1988), although we
have chosen to stress the positive aspects of the equal opportunity for
long-term research careers, regardless of PhD source.  Domen \& Thronson's
sample is biased in that it is restricted to researchers who settle at the
largest, degree-granting
institutes, while our unbiased sample includes
researchers at institutes large and small, foreign
and domestic, degree- and non-degree-granting.
\item ``Success'' is a dynamical entity - the survival analysis of Figure 1
demonstrates the danger of blindly adopting a 20-year baseline; in particular,
recent trends will be masked.
\item Within a decade of graduation, $\sim 45$\% of graduates will
have an identifiably permanent or tenured position, $\sim
20$\% remain in the soft-money/fixed-term contract category, and $\sim 35$\%
have left the field.
\item American graduates are far more likely to obtain permanent astronomy
positions in the USA ($\sim 34$\%), than are Dutch ($\sim 21$\%), Canadian
($\sim 17$\%), or Australian ($\sim 11$\%) graduates, in
their respective countries.
\item During the decade 1985-1994, the number of
astronomy PhDs granted by US, Canadian, and Dutch institutes 
increased by $\sim$20\% over the previous decade; in Australia, though, the
increase was $\sim$70\%.
\end{enumerate}

Again, if nothing else, the one point we wish to leave with the reader, or
prospective graduate student, is the optimistic primary conclusion above - i.e.,
regardless of the source of your PhD, one has just as good a chance to pursue a
research career in astronomy (albeit perhaps only at the part-time level), 
as the graduate of any other program.  It is true
that you may not have (statistically) the same odds as graduates of the
traditional ``prestigious'' schools in pursuing this career
at a large, degree-granting, university, but the research opportunities within
smaller universities and colleges (both 
degree and non-degree granting), overseas institutes, government labs, and
industry, appear to compensate.

\acknowledgements

The assistance of Don Faulkner, Gay Kennedy, John O'Byrne, and Tim de Zeeuw,
in compiling the MSSSO, Sydney, and Leiden statistics, is gratefully 
acknowledged.  We especially thank Virginia Trimble, for her careful reading of
the manuscript, and the subsequent lively exchange of extremely helpful email!
Finally, this paper may not have seen the light of day, without the aid of
Kevin Marvel at the AAS; special thanks go out to him.
This research has made \it extensive \rm
use of NASA's Astrophysics Data System Abstract Service,
the Science Citation Index, and the Institute of Scientific Information's
Citation Database.


\begin{references}
\reference Australian Astronomy: Beyond 2000, Prepared by the National Committee
for Astronomy of the Australian Academy of Science, Australian Government
Publishing Service, Canberra (June 1995)
\reference Domen, R.E. \& Thronson, Jr., H.A. 1988, \pasp, 100, 641
\reference Gourman, J. 1997, The Gourman Report, 8th Edition, National
Education Standards, Los Angeles
\reference Thronson, Jr., H.A. 1991, PASP, 103, 90
\reference Trimble, V. 1991, Scientometrics, 20, 71
\reference U.S. News Graduate School Rankings - 1998 
(Astrophysics/Space) -\hfil\break
\tt http://www.usnews.com/usnews/edu/beyond/\rm
\reference van der Kruit, P.C. 1994, Scientometrics, 31, 155
\end{references}
\end{document}